\documentclass[12pt,preprint]{aastex}

\begin{document}

\title{Novae as a Mechanism for Producing Cavities around
the Progenitors of SN~2002ic and Other SNe~Ia} 

\author{W. M. Wood-Vasey \and J. L. Sokoloski}

\date{{Accepted to ApJL:} May 24, 2006}

\begin{abstract}
We propose that a nova shell
ejected from a recurrent nova progenitor system
created the evacuated region around the explosion center of SN~2002ic.
In this picture, periodic shell ejections due to nova explosions on a
white dwarf sweep up the slow wind from the binary companion, creating
density variations and instabilities that lead to structure in the
circumstellar medium (CSM).  Our model naturally explains the observed gap between the
supernova explosion center and the CSM in SN~2002ic, accounts for the density
variations observed in the CSM, and resolves the coincidence problem
of the timing of the explosion of SN~2002ic with respect to the
apparent cessation of mass-loss in the progenitor system.  We also
consider such nova outburst sweeping as a generic feature of Type Ia
supernovae with recurrent nova progenitors.
\end{abstract}


\section{Introduction}

The remarkable supernova (SN) SN~2002ic has opened a window into the
nature and progenitors of Type Ia supernovae (SNeIa).  Discovered by
\citet{wood-vasey02} and identified from a pre-maximum spectrum as a
SNIa by \citet{hamuy02}, SN~2002ic appeared to be a normal SNIa 
from $\sim5$--$20$~days after explosion \citep[there were no 
observations during the first
$\sim5$~days after explosion;][]{wood-vasey04}.  
Around 22~days after explosion, SN~2002ic brightened to twice the
luminosity of a normal SNIa and showed strong H$\alpha$ emission
\citep{hamuy03a}.
The hydrogen emission and excessive brightness 
were attributed to interaction with a substantial circumstellar medium
\citep[CSM;][]{hamuy03b,deng04,wood-vasey04}.
The standard brightness before 22~days and the suddenness of
the brightening at that time implied that the original SN
explosion expanded into a region with little CSM,
i.e., a cavity,
and then suddenly encountered a region with significant CSM.
At 60~days after explosion, the lightcurve of SN~2002ic rose
again~\citep{hamuy03b,wood-vasey04} in a manner consistent with a
second increase in the CSM density.
SN~2005gj
appears to be another member of this class~\citep{prieto05}, 
and SN~1997cy is now widely believed to have
been a SN~2002ic-like event as well~\citep{hamuy03b,deng04,wood-vasey04}.

In the prevailing model for the production of SNeIa,  
a white dwarf accretes material from a companion star 
until the white dwarf approaches its Chandrasekhar
limit and explodes due to runaway thermonuclear
fusion~\citep[for recent reviews see][]{hillebrandt00, livio01,
wheeler02}.  In this so-called single-degenerate model,
the companion star can donate mass to the white dwarf
through a variety of mechanisms, 
including Roche-lobe overflow,
stellar winds, 
or expansion of the donor star to form a common envelope around both stars.  
Roche-lobe overflow results in a progenitor system whose environment is 
relatively free of hydrogen, while the
other two donor processes should expel hydrogen into the surrounding medium.
However, hydrogen had never been observed in a SNIa system prior
to the discovery of SN~2002ic.

The observation of significant amounts of hydrogen in SN~2002ic has
led to a flurry of speculation as to the nature of its progenitor
system.
SN~2002ic has been alternately modeled as the explosion of a white dwarf in a
binary system with a post-asymptotic giant branch (AGB)
companion~\citep{hamuy03b}; the explosion of the carbon-oxygen (C/O)
core of a 25~$M_\Sun$ star,  
a so-called SN~I\onehalf~event~\citep{iben83,hamuy03b,imshennik05}; the
merger of a white dwarf with the C/O core of an AGB star during a
common-envelope phase~\citep{livio03}; and the explosion of a white dwarf 
in a supersoft X-ray system with delayed dynamical instability-triggered
mass loss~\citep{han06}.
While these models each seek to explain the presence of a significant
amount of hydrogen
\citep[1--6$\,$~$M_\Sun$;][]{hamuy03b,chugai04,wang04} 
near the site of the thermonuclear explosion, none can easily account for
the size of the observed $\sim1.7\times10^{15}$~cm CSM-free region immediately
surrounding the explosion~\citep{wood-vasey04}.
In this paper we provide a mechanism that explains
this cavity as not due to the cessation of mass loss from the companion
star but rather due to the wind material having been swept up.
We propose that the progenitor of SN~2002ic was a white dwarf
undergoing recurrent novae while accreting from an AGB or post-AGB
companion and that the cavity surrounding the system was created by a
nova ejection approximately 15~years before the supernova explosion.

We describe our model for the cavity and CSM density variations around
SN~2002ic in Sec.~\ref{sec:2002ic}.  
In Sec.~\ref{sec:sneia} we discuss the cavities that could be produced
by prior novae in normal SNIa progenitor systems.
Finally, in Sec.~\ref{sec:discussion} we discuss our model in the
context of the diversity and asymmetry in SNIa progenitor systems and
make testable predictions.

\section{A Nova Cavity around SN~2002ic}
\label{sec:2002ic}

Unless material accreted onto the surface of a white
dwarf in an interacting binary burns quasi-steadily,
the white dwarf will experience thermonuclear flashes.  These nova
outbursts typically recur every $\sim10^5$~years~\citep{livio92}.
Immediately before
a SNIa explosion, however, the white dwarf mass is near the
Chandrasekhar limit, and the binary is likely to become a recurrent
nova (RN), in which outbursts recur every few years to decades
\citep[e.g.,][]{fuji82,prial82,starrfield85}.
About half of known RN are symbiotic RN, in which the white dwarf is
fed by the wind from a red giant.
The other half are generally close binary systems where the white
dwarf is accreting mass from its companion via Roche-lobe overflow.
Typical symbiotic RN consist of a white dwarf accreting from the wind
of a first-ascent red giant, but AGB star donors are also possible.
Such systems deposit hydrogen-rich material into their CSM 
through a wind from the red giant and possibly
also a hot wind from the white dwarf
\citep{hachisu99a,hachisu99b}.  The thermonuclear flashes
from RN eject mass shells with typical velocities of $v_\mathrm{shell} =
1000$--$4000$~km~s$^{-1}$ and masses of $M_\mathrm{shell} =
10^{-7}$--$10^{-5}$~$M_\Sun$~\citep{yaron05}.  In symbiotic RN, where the white dwarf is
embedded in the wind of the companion, the nova shell sweeps up this
wind and creates a cavity around the binary.

The significant CSM around SN~2002ic~\citep{hamuy03b} suggests that
the progenitor of that SNIa contained an extreme AGB star rather than
a first-ascent red giant companion.
We thus adopt a single-degenerate system consisting of a white dwarf
and a companion AGB star for our model of the progenitor system of
SN~2002ic.  
Our nova clearing mechanism, however, is not critically dependent 
on the choice of companion.
While early estimates of the total CSM
mass in the progenitor system suggested up to 6~$M_\Sun$ of circumstellar
material, proper treatment of the effect of the ejecta metallicity on
the X-ray absorption reduce the required CSM mass to $\sim1.1$~$M_\Sun$
within $3\times10^{16}$~cm of the progenitor system
\citep{chugai04,nomoto06b}.  The mass-loss rate for a progenitor
system with a wind velocity of $v_\mathrm{wind} = 10\,\mathrm{km}\,{\rm
s}^{-1}$ must thus have been close to $\dot{M}_\mathrm{wind} =
10^{-3}$~$M_\Sun$~yr$^{-1}$ in the thousand years before the SNIa explosion.
To model the progenitor of SN~2002ic, 
we thus take an AGB mass-loss rate of $\dot{M}_\mathrm{wind}=1\times10^{-3}$~$M_\Sun$~yr$^{-1}$~\citep{chugai04,nomoto06b},
a wind speed of $v_\mathrm{wind}=10$~km~s$^{-1}$~\citep{kudritzki78,gehrz71},
an ejected nova shell mass of $M_\mathrm{shell}=1\times10^{-6}$~$M_\Sun$, 
and
a shell velocity of $v_\mathrm{shell}=4000$~km~s$^{-1}$.  We assume that the white dwarf is
$\sim$10~AU from the companion star 
and that the CSM density falls off as $\rho(r_\mathrm{WD})
\propto 1 / r_\mathrm{WD}^2$ from the red-giant wind density at the
position of the white dwarf, where
$r_\mathrm{WD}$ is the distance from the white dwarf.  
Taking a nova recurrence time of $\sim20$~years, 
the accretion rate onto the white dwarf is $\gtrsim 5\times10^{-8}$, 
which is well within the recurrent nova mass accretion range of
$3\times10^{-8}$--$4\times10^{-7}$~$M_\Sun$~yr$^{-1}$
\citep{iben82,starrfield85,yaron05}.

A nova shell ejected into the dense material surrounding the white
dwarf progenitor of SN~2002ic produces a blast wave that quickly (in
minutes to hours) sweeps up enough CSM material to transition from
free expansion to deceleration in the Sedov-Taylor phase.
The CSM is so dense that the shock
becomes radiative $\sim5000$~seconds after the nova outburst and
enters the momentum-conserving snowplow phase.
A radiative shock moving in a $\rho \propto r^{-2}$
medium expands as $r_\mathrm{shock}\propto t^{1/2}$.  
While
\citet{nomoto06b} suggest that far from the progenitor white dwarf
the average CSM density falls off as $r^{-1.8}$,
which would lead to $r_\mathrm{shock}\propto t^{1/2.2}$, for
simplicity and maximal generality, we treat the CSM density as
following an overall $r^{-2}$ behavior.
Between the time of the last nova outburst and the SNIa explosion, 
the nova blast wave 
sweeps up the CSM and evacuates a cavity
with outer radius
\begin{equation}
R_\mathrm{out}
\approx R_\mathrm{SP}
\left(\frac{\Delta\,t_\mathrm{nova}}{t_\mathrm{SP}}\right)^{1/2},
\label{eq:outer_radius}
\end{equation}
where $R_\mathrm{SP}$ is the radius of the transition to the snowplow phase,
$t_\mathrm{SP}$ is the time between the nova explosion and the
onset of the snowplow phase of the nova blast wave, 
and $\Delta\,t_\mathrm{nova}$ is the time between 
the last nova outburst and the SNIa explosion.
While the nova blast wave expands and clears
out the surrounding CSM, the ongoing mass-loss from the system refills
a small portion of the cavity immediately surrounding the binary.

When the white dwarf explodes as a SNIa, the SN ejecta will first
encounter the mass lost from the AGB star since the last nova and
then the region that was evacuated by the nova blast wave.  Since the
observations of SN~2002ic began 5~days after the SN
explosion and no sign of CSM interaction was observed at that time,
the SN ejecta must have already moved through the small region
containing AGB wind material 
emitted since the last nova outburst.
\citet{gerardy04} calculated that up to $2\times10^{-2}$~$M_\Sun$ of CSM can be
overtaken by the SNIa ejecta at very early times with no appreciable
photometric signature.  
If the SN ejecta swept through any existing nearby CSM in the first 5
days after explosion, the last nova outburst likely occurred
within 
$\Delta\,t_\mathrm{nova} \le 5$~days$~(v_\mathrm{SN}/v_\mathrm{wind}) \approx14$~years 
of the SNIa explosion, where
$v_\mathrm{SN}=10^{4}$~km~s$^{-1}$ is the velocity of the SNIa
ejecta.  This $14$-year period is comparable to the typical
time between outbursts of recurrent novae.  If we take
$\Delta\,t_\mathrm{nova}=14$~years, the evacuated region has an outer
radius of $R(t_\mathrm{SN})\approx 1.5\times 10^{15}$~cm, 
extremely close to the cavity radius of 
$1.7\times10^{15}$~cm from the model of \citet{wood-vasey04}.  
See Table~\ref{tab:parameters} for a summary of these model parameters
and resulting cavity size.

Both the repeated pulses of recurrent nova events and
the instabilities at the shock fronts should create sizable variations
in the density structure of the CSM.
We interpret the secondary brightening of SN~2002ic around 60 days after the
SNIa explosion as due to a CSM density enhancement from a previous nova
outburst that occurred 50~years before the explosion of SN~2002ic.  
There are no
observations of SN~2002ic from 100--200~days after explosion, so 
no information is available about 
additional large-scale density 
variations from earlier nova outbursts.  In any event, as the system
evolves, the effects of earlier nova outbursts become smeared out as
hydrodynamic instabilities grow and the nova-driven shocks slow down
and approach $v_\mathrm{wind}$.
Hydrodynamically unstable interactions of the nova outburst shells with the stellar
wind outflow could produce the oft-discussed clumpiness 
in the CSM of SN~2002ic.

\section{Nova Cavities around Normal Type Ia Supernovae}
\label{sec:sneia}

The blast wave from a nova outburst will clear out the region around a
white dwarf in any recurrent nova SNIa progenitor.
Our model is most relevant for systems where one might otherwise
expect the presence of a significant amount of hydrogen.
Here we consider such a system consisting of a near-Chandrasekhar mass
white dwarf accreting material from a red giant companion at a
suitable rate to generate recurrent novae.
When the white dwarf experiences a nova outburst,
the blast wave expands freely for a few days, 
then experiences a several-month Sedov-Taylor phase, 
and finally enters the momentum-conserving snowplow phase. 
Observations of the recurrent nova RS~Oph 
suggest $v_\mathrm{shell} \approx 4000$~km~s$^{-1}$~\citep{bode87}, a 2-day
free-expansion phase \citep{sokoloski06}, and a 2-month Sedov-Taylor
phase~\citep{mason87}.  Taking a typical symbiotic-star mass-loss rate
of $\dot{M}_\mathrm{wind} \sim 10^{-7}$~$M_\Sun$~yr$^{-1}$~\citep{seaquist90}, an ejected shell mass
of $M_\mathrm{shell}=4\times10^{-7}$~$M_\Sun$~\citep{hachisu01}, an
ejected shell velocity of $v_\mathrm{shell}=4000$~km~s$^{-1}$, and a
binary separation of 1~AU, the nova blast wave would evacuate a
region with radius $\sim9.9\times10^{15}$~cm in 40~years.  During this
time, the red-giant wind would refill the innermost
volume, out to $\sim1.3\times10^{15}$~cm.  It would take
SNIa ejecta traveling at $10^{4}$~km~s$^{-1}$ $\sim15$~days to
traverse the refilled region and
$\sim115$~days to reach the outer
edge of the cavity.  
See Table~\ref{tab:parameters} for a summary of these model quantities.

\begin{deluxetable}{lrcccccll}
\tablecolumns{9}
\tablewidth{0pc}
\tablecaption{Model parameters}
\tablehead{
\colhead{Model} & \colhead{Binary}     & \colhead{$\dot{M}_\mathrm{wind}$} & \colhead{$M_\mathrm{shell}$} & \colhead{$v_\mathrm{shell}$} & \colhead{$v_\mathrm{SN}$} & \colhead{$\Delta\,t_\mathrm{nova}$} & \multicolumn{2}{c}{Cavity Extent} \\
\colhead{}      & \colhead{Separation} & \colhead{}                        & \colhead{}                   & \colhead{}                   & \colhead{}                & \colhead{}                          & \colhead{Inner}        & \colhead{Outer}           \\
\colhead{}      & \colhead{[AU]}       & \colhead{[$M_\Sun$/year]}         & \colhead{[$M_\Sun$]}         & \colhead{[km/s]}             & \colhead{[km/s]}          & \colhead{[years]}                   & \multicolumn{2}{c}{[$10^{15}$ cm]}     \\
}                                                                                                                                                                    
\startdata                                                                                                                                                           
SN~2002ic       &      10              &            $10^{-3}$              &   $1\times 10^{-6}$          &   4000                       &       $10^{4}$            &        14                           &  0.44 & 1.5      \\
WD+RG $\Rightarrow$ SNIa      &       1              &            $10^{-7}$              &   $4\times 10^{-7}$          &   4000                       &       $10^{4}$            &        40                           &  1.3 & 9.9      \\
\enddata
\tablecomments{Model parameters for the nova outburst clearing mechanism for the case of SN~2002ic and for a more generic SNIa progenitor system consisting of a white dwarf with a red giant companion.  See text for more detail on the parameter definitions.}
\label{tab:parameters}
\end{deluxetable}

\section{Discussion}
\label{sec:discussion}

We propose that recurrent nova outbursts can clear out hydrogen-free
regions around the progenitor systems of SNeIa and thus explain the
general lack of observed hydrogen in SNeIa as well as the
clearly-defined hydrogen-free region around SN~2002ic.  Our model
provides an explanation for 
what had previously been seen as a coincidence in the timing of the 
end of mass-loss in the progenitor system 
and the explosion of SN~2002ic.
The size of the cavity cleared out in SN~2002ic-like events is
determined by the time between the last nova outburst and the SN
explosion.
We attribute the relative rarity of SN~2002ic-like events to the
scarcity of binary systems with such extreme mass-loss rates.

\citet{wang04} and \citet{deng04} both discuss a model for the CSM
of SN~2002ic that involves a significant asymmetry in the form of a
clumpy disk surrounding the progenitor system.  The arguments for this
interpretation derive both from spectropolarimetry~\citep{wang04} and
the late-time observations of high velocities
($\sim10^{4}$~km~s$^{-1}$) that are inconsistent with ejecta having
been slowed down by interaction with the CSM \citep{deng04}.
Nova outbursts are also observed to be
asymmetric~\citep[e.g.,][]{taylor89,anupama94,rupen06}.  
While the nova outburst-CSM interaction becomes somewhat more
complicated in this more complex geometry, the basic principle
remains, and the nova blast wave still clears out substantial cavities
around the SNIa progenitor.
In addition, our model naturally generates large-scale structure that
could lead to observed clumpiness in such disks.

Since any near-Chandrasekhar-mass white dwarf accreting at a rate
between $\sim10^{-8}$ and $4 \times 10^{-7}$~$M_\Sun$~yr$^{-1}$ will experience
recurrent nova outbursts~\citep{iben82,starrfield85,yaron05}, the
effect of a nova blast wave on the CSM must be taken into account when
interpreting the limits placed by observations on the mass-loss rate
of SNIa progenitors.  Apparent upper limits on mass-loss rates from
the progenitors of SNeIa (assuming a wind speed of
$v_\mathrm{wind}=10$~km~s$^{-1}$) include $2$--$3\times10^{-4}$~$M_\Sun$~yr$^{-1}$
\citep[SN~1992A;][]{schlegel93}; $1\times10^{-7}$~$M_\Sun$~yr$^{-1}$
\citep[SN~1986G;][]{eck95}; and $9\times10^{-6}$~$M_\Sun$~yr$^{-1}$ \citep[from
the unusual SN~2000cx;][]{lundqvist03}.  Recent work by
\citet{panagia06} from VLA observations of 27~SNeIa resulted in more
a more constraining upper limit of $3\times10^{-8}$~$M_\Sun$~yr$^{-1}$
(assuming a steady outflow with $v_\mathrm{wind}=10$~km~s$^{-1}$).
These findings would appear to rule out many classes of
single-degenerate progenitors.
However, if the action of prior novae is taken into account, 
typical mass-loss rates for single-degenerate SNIa progenitor models
are easily
compatible with the aforementioned observations.  
In Sec.~\ref{sec:sneia} we considered an example system with 
$\dot{M}_\mathrm{wind}=1\times10^{-7}$~$M_\Sun$~yr$^{-1}$
that would lead to early and late-time emission that would not
have been seen by the observations of \citet{panagia06}.
Even higher mass-loss rates of up to $\sim10^{-6}$~$M_\Sun$~yr$^{-1}$ are
allowed if correspondingly more massive shells are ejected in the
nova outbursts.

The radius of the cleared-out cavity is a function of the
mass-loss rate of the progenitor system.
The different evolutionary stages of the companion stars to the white dwarf in
SNIa progenitor systems lead to qualitatively different wind mass-loss rates.
These different companion stars (e.g., first-ascent red giant vs. AGB)
can explain the distinctly different behavior of normal SNeIa and
SN~2002ic-like events.
 
We encourage 
radiative-hydrodynamic modeling 
to properly explore the effect of multiple nova outbursts on the
density structure of the surrounding CSM.  Our model suggests that
SN ejecta-CSM interaction should be observable at late times for
normal SNeIa but also that the strength of the observed signal from such
interaction would be much lower than that for SN~2002ic.
The interaction of the SN ejecta with the wind blown since the last
nova outburst can also produce an observable signal at early times.
While such early interaction may not significantly affect the observed
lightcurve due to the limited amount of CSM involved,
\citet{gerardy04} and \citet{mazzali05} suggest that high-velocity
spectroscopic features would be signs of SN ejecta-CSM interaction at
early times.  We urge the observation of SNeIa at both very early and
late times to look for high-velocity features, H$\alpha$ emission, and
deviations from the expected exponential decay powered by $^{56}$Co.
We suggest that interaction of the SN ejecta with the CSM will give
rise to such early- and late-time features.

\section{Acknowledgments}

We thank our anonymous referee for helpful comments that improved the
clarity of this paper. 
WMWV and JLS are respectively supported by awards AST-0443378 and
AST-0302055 from the US National Science Foundation.

\end{document}